\documentclass{ifacconf}

\usepackage{graphicx}      
\usepackage{natbib}        
      
\usepackage{xcolor}
\usepackage{amsmath}
\usepackage{colortbl}
\usepackage{multirow}
\usepackage{booktabs}
\usepackage{bm}
\usepackage{tikzscale}
\usepackage{pgfplots}
\pgfplotsset{compat=1.18}
\usepackage{tikz}
\usepackage{amssymb}
\usepackage{accents}
\usepackage{mathtools} 

\newtheorem{assumption}{\textbf{Assumption}}
\newtheorem{proposition}{\textbf{Proposition}}

\newtheorem{corollary}{\textbf{Corollary}}
\newtheorem{theorem}{\textbf{Theorem}}

\newtheorem{remark}{\textbf{Remark}}

\newtheorem{definition}{\textbf{Definition}}

\definecolor{am}{rgb}{0.0, 0.6, 0.5}
\definecolor{cblue}{rgb}{0.16, 0.32, 0.75}
\definecolor{vio}{rgb}{0.83, 0.08, 0.52}
\definecolor{red}{rgb}{0.9, 0.17, 0.31}

\def\vecc{{\rm vec}}
\def\row{{\rm row}}
\def\cov{{\rm cov}}

\def\E{\mathbb{E}}

\def\R{\mathbb{R}}

\def\endproof{\begin{flushright} \vspace{-0.5cm} $\blacksquare$ \end{flushright}}

\newcommand{\X}{\ensuremath{\textcolor{am}{\bm{X}}}}
\newcommand{\SM}{\ensuremath{\textcolor{am}{\bm{S}}}}
\newcommand{\T}{\ensuremath{\textcolor{am}{\bm{T}}}}
\newcommand{\Az}{\ensuremath{\textcolor{am}{\bm{A_0}}}}
\newcommand{\Cz}{\ensuremath{\textcolor{am}{\bm{C_0}}}}
\newcommand{\Nz}{\ensuremath{\textcolor{am}{\bm{N_0}}}}
\newcommand{\NzT}{\ensuremath{\textcolor{am}{\bm{N_0^T}}}}
\newcommand{\Xii}{\ensuremath{\textcolor{am}{\bm{X^{(i)}}}}}

\begin{document}
\begin{frontmatter}

\title{Covariance Stabilization for a class of Stochastic Discrete-time Linear Systems using the S-Variable Approach\thanksref{footnoteinfo}} 

\thanks[footnoteinfo]{This work was partially supported by the Clinical project, funded by the ANR under grant ANR-24-CE45-4255. 
}

\author[UPHF,INSA]{Kaouther Moussa} 
\author[LAAS]{Dimitri Peaucelle} 

\address[UPHF]{UPHF, CNRS, UMR 8201 - LAMIH, F-59313 Valenciennes, France}                                              
 \address[INSA]{INSA Hauts-de-France, F-59313, Valenciennes, France}           

 \address[LAAS]{LAAS-CNRS, Univ. Toulouse, CNRS, Toulouse, France}

\begin{abstract}                
This paper deals with the problem of covariance stabilization for a class of linear stochastic discrete-time systems in the Stochastic Model Predictive Control (SMPC) framework. The considered systems are affected by independent and identically distributed (i.i.d.) additive and parametric stochastic uncertainties (potentially unbounded), in addition to polytopic deterministic uncertainties bounding the mean of the state and input parameters. The design conditions presented in this paper are formulated as Linear Matrix Inequalities (LMIs), using the S-variable approach in order to reduce the potential conservatism. These conditions are derived using a deterministic exact characterization of the covariance dynamics, the latter involves bilinear terms in the control gain. A technique to linearize such dynamics is presented, it results in a descriptor representation allowing to derive sufficient conditions for the design of a covariance-stabilizing controller. The derived condition is first compared with a known necessary and sufficient stability condition for systems without deterministic uncertainties and additive stochastic noise. Although more conservative, the proposed condition is more numerically tractable, with an LMI size scaling as \(\mathcal{O}(n^2)\) instead of \(\mathcal{O}(n^3)\). Then, the same condition is used to design controllers that are robust to both deterministic and stochastic uncertainties.  Several numerical examples are presented for comparison and illustration. 
\end{abstract}


\end{frontmatter}

\section{Introduction}
Stochastic systems allow to describe dynamical systems that are affected by uncertainties of probabilistic nature. Such a property is inherently present in various systems, for instance those modeling biological phenomena (see  \cite{DelVecchio2018} and references therein). This paper addresses the problem of stabilizing the covariance of a class of linear stochastic discrete-time systems in the Stochastic Model Predictive (SMPC) framework. The latter has gained a considerable interest in the last decade because of its ability to reduce the conservatism related to worst-case scenario-based approaches (see \cite{Mesbah2016} and references therein). 

One of the main techniques to address uncertainties (both deterministic and stochastic) in MPC methods is the tube-based one \cite{Langson}, it consists in decomposing the state into a deterministic and  an uncertain component, this decomposition allows to rewrite the SMPC problem as a deterministic one, which considerably simplifies the resolution of related optimization problems. In this context,  covariance dynamics characterization is useful for tightening time-varying chance constraints using concentration inequalities, such as the Chebyshev's inequality as used for example in \cite{KofmanAUT12,Hewing2018,Fiacchini2021}. Whenever a characterization of covariance dynamics is available, probabilistic constraints can be tightened using analytical formulations instead of sampling-based methods that were proposed, for example, in \cite{Cannon2011,calafiore2012,Lorenzen2016}. 

As mentioned in \cite{Moussa2025}, the characterization of covariance dynamics in SMPC has so far been limited to linear discrete-time systems affected by an additive noise. In this case, the derivation of the covariance dynamics is trivial and its stability is conditioned by the Schur stability of the closed-loop nominal system $\left(A+BK \right)$. In \cite{Moussa2025}, we proposed a new covariance dynamics characterization for systems with  stochastic parametric uncertainties in the state matrix. Since this characterization is deterministic, classical LMI-based techniques can be applied to derive synthesis conditions ensuring the covariance stability. The main challenge is the presence of bilinear terms in the control gain, due to a Kronecker product related transformation.  

In this paper, we address the class of stochastic discrete-time systems that are affected by parametric and additive stochastic uncertainties, in addition to polytopic deterministic uncertainties bounding the mean of the state and input parameters. Practically, such systems can describe the error that might occur when modeling the parameters with probability distributions, and allow to take into account deterministic uncertainties on the mean of the parameters. 

Similar problems have been addressed in the stochastic control literature, in particular the random polytopes framework presented for example in \cite{HOSOE2018,HOSOE2020} where an LMI-based robust stability condition was provided for systems described by random polytopes, whose vertices are random matrices with given distributions. Such conditions involve expectation-based inequalities and a linearization technique was proposed in \cite{HOSOE2020}  to derive deterministic sufficient conditions for the robust stability of such systems. 

The work in  \cite{HOSOE2020} was based on the work in \cite{HOSOE2019}, in which the authors proposed a necessary and sufficient condition for the stability of linear stochastic discrete-time systems, with i.i.d. parametric uncertainties and without the polytopic representation. The latter is closely related to our work and will be described in Section~\ref{Section:background}. The class of systems considered in this paper can therefore be seen as a special case of random polytopes, with a simpler representation allowing to derive more tractable LMI conditions. 

The contributions of this paper are threefold: deriving an LMI-based condition for stabilizing the covariance dynamics introduced in~\cite{Moussa2025}; assessing its conservatism and numerical tractability with respect to known necessary and sufficient conditions; and extending it to the case involving both deterministic polytopic and stochastic uncertainties.

The paper is organized as follows. 
Section~\ref{Section:Problem} states the problem and assumptions. 
Section~\ref{Section:background} recalls the necessary and sufficient stability condition of~\cite{HOSOE2019}. 
Section~\ref{Section:Cov_char} presents the covariance characterization and its extension to polytopic uncertainties. 
Section~\ref{Section:LMI} derives the proposed stabilization conditions. 
Section~\ref{Section:Num_tests} evaluates their tractability and conservatism and illustrates the polytopic case. 
Section~\ref{Section:Conclusion} concludes the paper and outlines future research directions.

\subsection{Notation}
The sets of real and natural numbers are denoted by \(\mathbb{R}\) and \(\mathbb{N}\), respectively. 
\(\mathbb{R}^n\), \(\mathbb{R}^{n\times m}\), \(\mathbb{S}^n\), and \(\mathbb{S}^n_+\) denote, respectively, the sets of real \(n\)-vectors, real \(n\times m\) matrices, symmetric matrices, and symmetric positive definite matrices. 
The expectation and covariance of a random vector \(v\) are denoted by \(\mathbb{E}[v]\) and \(\cov(v)\). 
The Gaussian distribution with mean \(\mu\) and covariance \(\Sigma\) is denoted by \(\mathcal{N}(\mu,\Sigma)\), and \(\otimes\) denotes the Kronecker product. 

For a square matrix \(A\), \(A^{-1}\), \(\mathrm{tr}(A)\), \(\rho(A)\), and \(\mathrm{mspec}(A)\) denote its inverse, trace, spectral radius, and eigenvalue multiset, respectively. 
A matrix is Schur stable if all its eigenvalues lie in the open unit disk. 
The zero and identity matrices of appropriate dimensions are denoted by \(0\) and \(I_n\). For \(B\in\mathbb{R}^{n\times m}\), \(\vecc(B)\) denotes its vectorization in a column-major order, with inverse operation \(\vecc^{-1}(\cdot)\). The notation $\row(B)$ stands for the row-major order stacking operation. The transpoe of $B$ is denoted $B^T$. 
Moreover, for a square matrix \(Q\),  \(\{Q\}^{\mathcal{S}}=Q+Q^T\). 
 The shorthand \((\star)QB\) denotes the symmetric expression \(B^TQB\). 
For symmetric matrices, \(M\succ0\) and \(M\succeq0\) denote positive definiteness and positive semidefiniteness, while \(M_1\prec M_2\) means \(M_2-M_1\succ0\). 
An inequality \(I(\X)\prec0\) is an LMI if \(I(\X)\) is affine in the decision variables \(\X\). 
Decision variables are highlighted in \textcolor{am}{\textbf{bold green}}.

\section{Problem statement}\label{Section:Problem}

Consider the following class of stochastic discrete-time linear systems:
\begin{equation}
    x_{k+1} = \tilde{A}(\xi_k, \theta) x_k + B(\theta) u_k + w_k.
\label{Eq:sys_dyn}
\end{equation} 
where \(x_k\in\mathbb{R}^n\), \(u_k\in\mathbb{R}^m\), and \(w_k\in\mathbb{R}^n\) is a zero-mean additive noise with covariance \(W\succ0\). The initial condition \(x_0\) is deterministic. The state matrix is decomposed as
\[
\tilde A(\xi_k,\theta)=A(\theta)+\bar A(\xi_k),
\]
where
\[
A(\theta)=\sum_{i=1}^{L}\theta^{(i)}A^{(i)},\qquad
B(\theta)=\sum_{i=1}^{L}\theta^{(i)}B^{(i)}.
\]
The deterministic uncertain vector $\theta = \left( \theta^{(i)}, \ldots,  \theta^{(L)}\right)^T$ satisfies $\theta \in \mathbf{E}^{L}$ with
\begin{equation*}
  \mathbf{E}^{L} \coloneqq 
  \left\{\theta\in\mathbb{R}^{L}:\theta^{(i)}\geq0,\ i=1,\ldots,L,\ 
  \sum_{i=1}^{L}\theta^{(i)}=1\right\},
\end{equation*}
which is assumed to contain the true value \(\theta^{\mathrm{true}}\). The pair \((A(\theta),B(\theta))\) is assumed to be robustly stabilizable.

The process \(\xi=(\xi_k)_{k\in\mathbb{N}}\) represents stochastic parametric uncertainties affecting the state matrix. Its support is \(\Xi\subset\mathbb{R}^l\), and \(\bar A:\Xi\to\mathbb{R}^{n\times n}\) satisfies \(\mathbb{E}[\bar A(\xi_k)]=0\). Hence, \(A(\theta)=\mathbb{E}[\tilde A(\xi_k,\theta)]\).

In practice, the class of systems represented by (\ref{Eq:sys_dyn}) encompasses systems subject to stochastic uncertainties in the state parameters, whose mean values are themselves uncertain yet bounded within a polytope.

In this paper, the stochastic processes \(w\) and \(\xi\) are considered to satisfy the following assumption.

\begin{assumption}\label{Ass:iid}
For any stochastic process \(\nu=(\nu_k)_{k\in\mathbb{N}}\), \(\nu_k\) is independent and identically distributed with respect to \(k\).
\end{assumption}

The following integrability assumption ensures that the second-order moments of the entries of \(\bar A(\xi_k)\) are well defined.

\begin{assumption}\label{Ass:Leb_integr}
The entries of \(\bar A(\xi_k)\) are square integrable, i.e.,
\[
\mathbb{E}\!\left[\bar A_{ij}(\xi_k)^2\right]<\infty,\qquad i,j=1,\ldots,n .
\]
\end{assumption}

Since \(\xi_k\) and \(w_k\) are i.i.d. in time, \(\bar A(\xi_k)\) and \(w_k\) are independent of \(x_k\) and \(u_k\). Moreover, \(w\) is assumed independent of \(\xi\).

\begin{assumption}\label{Ass:ind_w_AB}
The process \(w\) is independent of \(\xi\), i.e., for all \(k\in\mathbb{N}\),
\[
\mathbb{E}[\xi_{ik}w_{jk}]
=
\mathbb{E}[\xi_{ik}]\mathbb{E}[w_{jk}]
=0,
\quad i=1,\ldots,l,\; j=1,\ldots,n .
\]
\end{assumption}

Assumption~\ref{Ass:ind_w_AB} is only used to simplify the covariance characterization; possible additional terms arising otherwise do not affect the subsequent stability analysis.

The goal is first to derive stabilizing design conditions, in the tube-based SMPC framework, for the covariance dynamics of the particular case
\begin{equation}
    x_{k+1} = \tilde{A}(\xi_k) x_k + B u_k + w_k,
\label{Eq:sys_dyn_A}
\end{equation}
where \(\theta^{\text{true}}\) is known, so that \(A(\theta^{\text{true}})=A\) and \(B(\theta^{\text{true}})=B\). 
The resulting condition is then extended to system~(\ref{Eq:sys_dyn}).

\section{Related background}\label{Section:background}
Consider the stochastic linear system
\begin{equation}
    x_{k+1}=\tilde A(\xi_k)x_k+\tilde B(\xi_k)u_k ,
\label{Eq:sys_dyn_no_w}
\end{equation}
where the entries of \(\tilde A(\xi_k)\) and \(\tilde B(\xi_k)\) are square integrable. 
For systems satisfying Assumptions~\ref{Ass:iid} and~\ref{Ass:Leb_integr}, \cite{HOSOE2019} proved the equivalence between asymptotic, exponential, and quadratic mean-square stability. We recall the following definition.

\begin{definition}\label{Def:asy_stab}
System \(x_{k+1}=\tilde A(\xi_k)x_k\) is mean-square asymptotically stable if, for each \(\varepsilon>0\), there exists \(\delta(\varepsilon)>0\) such that
\[
\|x_0\|\leq\delta(\varepsilon)
\Rightarrow
\mathbb{E}[\|x_k\|^2]\leq\varepsilon,\quad \forall k\in\mathbb{N},
\]
and \(\mathbb{E}[\|x_k\|^2]\to0\) as \(k\to\infty\), for all deterministic \(x_0\in\mathbb{R}^n\).
\end{definition}

For the uncontrolled system, quadratic stability is equivalent to the existence of \(P\in\mathbb{S}^n_+\) such that
\[
\mathbb{E}\!\left[P-\tilde A(\xi_0)^T P\tilde A(\xi_0)\right]\succ0 .
\]
For system~(\ref{Eq:sys_dyn_no_w}) with \(u_k=Fx_k\), \cite{HOSOE2019} derives a necessary and sufficient LMI condition by decomposing the matrix
\begin{equation}
    \E \left[(\star)\left[\row \left(\tilde{A}(\xi_0)\right),\: \row \left( \tilde{B}(\xi_0) \right) \right] \right].
    \label{Eq:expectation_matrix}
\end{equation}
The resulting condition is recalled below.
\begin{theorem}\cite{HOSOE2019}
    Let $\xi$  satisfy Assumption~\ref{Ass:iid}, and assume that the square elements of $\tilde{A}(\xi_k)$ and $\tilde{B}(\xi_k)$ are Lebesgue integrable. There exists a gain $F$ such that the closed-loop system defined by $x_{k+1} = \left(\tilde{A}(\xi_k) +  \tilde{B}(\xi_k) F\right) x_k$ is quadratically stable if and only if there exist $X \in \mathbb{S}^{n}_+$ and $Y \in \mathbb{R}^{m \times n}$ satisfying:
    \begin{equation}\label{LMI:Yohei2019}
          \begin{pmatrix}
        X & * \\
        \bar{G}_A^{'} X + \bar{G}_B^{'} Y & X \otimes I_{\bar{n}}
    \end{pmatrix} \succ 0,  
    \end{equation}
  where $\bar{G}_A^{'} \in \R^{n \bar{n} \times n}$ and $\bar{G}_B^{'}\in \R^{n \bar{n} \times m}$ result from the expectation-based decomposition of the matrix in (\ref{Eq:expectation_matrix}) as presented in \cite{HOSOE2019}, with $\bar{n} \leq n(n+m)$ as suggested in \cite{Hosoe2019b}, and $F = Y X^{-1}$ is such a stabilizing gain.
    \label{Th:Yohei2019}
\end{theorem}
The reader is referred to~\cite{HOSOE2019} for the decomposition details and decay-rate extensions. 
The LMI in Theorem~\ref{Th:Yohei2019} has size at most \(n+n^2(n+m)\). 
When the matrix in~(\ref{Eq:expectation_matrix}) is rank-deficient, e.g., when either the state or input matrix is deterministic, this size can be reduced by removing the part associated with zero singular values. 
This reduction is used in Section~\ref{Section:Num_tests} for a fair comparison.

\section{Exact covariance characterization in the SMPC framework}\label{Section:Cov_char}
Tube-based decomposition is one of the main approaches to handle uncertainties in robust and stochastic MPC frameworks, see for example \cite{Langson} and \cite{Arcari2023}. It consists in separating the state into a deterministic and an uncertain component, and designing a pre-stabilizing feedback that allows to cope with the uncertain state. In the framework of stochastic MPC, the tube-based approach allows to obtain a tractable formulation that can handle the presence of chance constraints. Therefore, characterizing the covariance dynamics is crucial in this context. 

This section uses our previous work in \cite{Moussa2025} on the exact  characterization of the covariance dynamics related to the class of systems in~(\ref{Eq:sys_dyn_A}), to extend it to the case of system~(\ref{Eq:sys_dyn}).

As commonly done in tube-based MPC approaches, we consider the following state decomposition:
\begin{equation}
    x_k = e_k + z_k,
\end{equation}
where $e_k$ stands for the uncertain state (also denoted as the error), and $z_k$ for the deterministic part representing the mean of the state $x_k$ that we define as follows:
\begin{equation}
    z_{k+1} = A(\theta) z_k + B(\theta) v_k
\end{equation} 
with $z_0 = x_0$ and then $e_0=0$. We restrict the class of control policies over the following affine error feedback law:
\begin{equation}
    u_k = K e_k + v_k, 
\end{equation}
with $K \in \R^{m \times n} $.  Then, the resulting error dynamics related to $e_k$ can be written as follows:
\begin{equation}
    e_{k+1} = \left(\tilde{A}(\xi_k,\theta)+B(\theta)K \right) e_k + \bar{A}(\xi_k) z_k + w_k
\end{equation}
The following standard assumption is functional to the subsequent results and is not restrictive since $(A(\theta),B(\theta))$ is assumed to be robustly stabilizable, as commonly considered in standard robust MPC methods \cite{LORENZEN}.

\begin{assumption}
The control input $v_k$, defined by a state feedback law $v_k = \bar{K}(z_k)$ is designed such that the closed-loop system $z_{k+1}=\left(A(\theta)z_k + B (\theta)\bar{K}(z_k) \right)$ is asymptotically stable $\forall \: \theta \in \mathbf{E}^{L}$. 
\label{Ass:vk_stab}
\end{assumption}

In the case of control without constraints, this is a classical robust state-feedback problem that can be solved with LMI techniques (see for example in \cite{Ebihara2015}). In the case of a tube-based stochastic MPC implementation, $v_k$ can be designed by a deterministic MPC strategy \cite{Arcari2023}.  

The following proposition gives the error covariance dynamics of system~(\ref{Eq:sys_dyn}) which is a direct extension of Theorem~1 in \cite{Moussa2025}. The reader is referred to that work for the detailed proof. 
\begin{proposition} \label{Pr:exact_char} 
The dynamics of the error covariance related to system~(\ref{Eq:sys_dyn}) is given by:  
\begin{align}
\cov(e_{k+1})&=
(\star)\cov(e_k)\left(A(\theta)+B(\theta)K\right)^T + W   \nonumber\\
&\hspace{-1cm} + \vecc^{-1} \left(\mathbb{E}\left[\bar{A}(\xi_k)\otimes\bar{A}(\xi_k)\right]\vecc\left( \cov(e_k)+z_kz_k^T\right)\right)\nonumber
\end{align}
which equivalently reads as 
\begin{align}
    \epsilon_{k+1}&= M(\theta,K) \epsilon_k 
    + C_p^A \zeta_k + \omega
\label{eq:err_cov_dynamics}
\end{align}
where $\epsilon_k=\vecc\left(\cov(e_k) \right)$, $\zeta_k = \vecc \left( z_kz_k^T\right)$, $\omega = \vecc \left( W\right)$,  $C_p^A=\mathbb{E}\left[\bar{A}(\xi_k)\otimes\bar{A}(\xi_k)\right]$ and
\[M(\theta,K)=\Big( \left(A(\theta)+B(\theta)K\right) \otimes \left(A(\theta)+B(\theta)K\right) + C_p^A \Big).\]\nonumber
\end{proposition} 
Note that the resulting matrix $C_p^A$ is constant and contains the different second-order moments related to the parameters joint distribution, it can be seen as a transformation of the covariance matrix related to the state parameters. Therefore, this proposition presents a deterministic characterization of the error covariance  dynamics for system~(\ref{Eq:sys_dyn}).

 
 The following proposition establishes a sufficient condition for the stability of the covariance dynamics in (\ref{eq:err_cov_dynamics}), which is also an extension of the condition presented in \cite{Moussa2025} without polytopic uncertainties.
 \begin{proposition} 
If $\bar K$ stabilizes robustly the deterministic system as in Assumption~\ref{Ass:vk_stab} and $K$ robustly  stabilizes the deterministic system $\epsilon_{k+1}=M(\theta,K)\epsilon_k$ then the covariance matrix $\cov(e_k)$ converges to \begin{equation}
\vecc^{-1}\left(\left( I-M(\theta^{\text{true}},K)\right)^{-1} \vecc\left(W\right)\right).
\end{equation}  
\end{proposition}
The proof follows from the steady-state dynamics of (\ref{eq:err_cov_dynamics}), under the satisfaction of Assumption~\ref{Ass:vk_stab}.


\section{LMI-based control design conditions}\label{Section:LMI}

While finding robust state-feedback gains $\bar K$ for linear uncertain systems with matrices lying in polytopes is a well established topic, we now consider the more involved problem of finding $K$ such that the bilinear matrix $M(\theta,K)$ is robustly Schur stable. Before moving to the robust case, we assume $\theta=\theta^{\text{true}}$ is known and set $M(\theta^{\text{true}},K)=M(K)$, 
$A(\theta^{\text{true}})=A
$, $B(\theta^{\text{true}})=B$. 

The difficulty of $M(K)$ being bilinear is circumvented by descriptor modeling and matrix inequalities are obtained using the S-variable approach.




\begin{theorem}\label{TH:stab_cond_A}
Assume there exists $\X \in \mathbb{S}^{n^2}_+$, $\SM\in \R^{n\times n}$, $\T \in \R^{m \times n}$ and
$\Nz \in \R^{3n^2\times 2n^2}$
the following holds: 
\begin{equation}
    \begin{pmatrix} 
    \X& 0 & 0  & \\
 0 & - \X & 0  & \\
  0 & 0 & 0  & 
\end{pmatrix} \prec 
\left\{ N(\SM,\T)\NzT
\right\}^{\mathcal{S}}
\label{LMI:specific}
\end{equation}  
where
\[
N(\SM,\T)=
\left(\begin{array}{cc|  cc }
- \SM \otimes I_n && 0  \\
C_p^A\left(\SM\otimes I_n \right) && I_n \otimes (A \SM+B\T) \\
(A\SM+B\T)\otimes I_n && -I_n\otimes \SM   \\
\end{array}\right)
\]
then $K = \T \SM^{-1}$ is such that $M(K)$ 
is Schur stable. 
\end{theorem}
\paragraph*{Proof}
Consider the dual system $\alpha^T_{k+1}=\alpha^T_k M(K)$ which is Schur stable if and only if $\epsilon_{k+1}=M(K)\epsilon_k$ is Schur stable. Let $\pi_k^T=\alpha_k^T\left(I_n\otimes (A+BK)\right)$ and $\eta_k^T=\left(\alpha_{k+1}^T,\alpha_k^T,\pi_k^T\right)$. The dual system dynamics also read as the following affine in $K$ descriptor form:
\begin{align}
\eta_k^T
\left(
\begin{array}{cc|  cc }
- I_{n^2} && 0 \\
C_p^A && I_n \otimes (A+BK)\\
(A+BK)\otimes I_n && - I_{n^2} \nonumber
\end{array}\right)  =0.  
\end{align}   
Post multiplying this equality by 
\[
\begin{pmatrix}
   \SM \otimes I_n & 0  \\
    0 &  I_n \otimes \SM\nonumber
\end{pmatrix}
\]
together with $\T=K\SM$ gives that the following holds along the trajectories
\(
\eta_k^T N(\SM,\T)=0.
\)
Therefore (\ref{LMI:specific}) implies that 
\[
\eta_k^T\begin{pmatrix} 
    \X& 0 & 0  & \\
 0 & - \X & 0  & \\
  0 & 0 & 0  & 
\end{pmatrix}\eta_k
=
\alpha_{k+1}^T \X \alpha_{k+1}-\alpha_k^T \X \alpha_k
<0
\]
hold for the system. The matrix $\X$ is hence a Lyapunov certificate that proves stability.

\endproof

\begin{remark}
Note that the dimension of the condition in~(\ref{LMI:specific}) is $3n^2$ which is, in general, lower than the dimension of the necessary and sufficient condition in (\ref{LMI:Yohei2019}). 
\end{remark}

\subsection{Comments on the choice of $N_0$}

The above proof is valid whatever matrix $\Nz$ as long as (\ref{LMI:specific}) holds. Moreover, for a fixed value of $\Nz$ the conditions are LMIs in the decision variables $\X$, $\SM$ and $\T$. It is hence of crucial importance to have clues on how to chose the $\Nz$, which can be critical considering the large size of the matrix.

According to results in \cite{Ebihara2015}, the following structure is non conservative 
\[
\Nz=
\left(\begin{array}{cc|  cc }
- I_{n^2} && 0  \\
\Cz && I_n \otimes \Az \\
\Az\otimes I_n && -I_{n^2}
\end{array}\right)
\]
More precisely, using Finsler's lemma one can prove that if the exists $K$ solution to the problem then there necessarily exist $\Az=A+BK$ and $\Cz=C_p^A$ solution to (\ref{LMI:specific}).

Moreover, for (\ref{LMI:specific}) to hold, it is necessary that the system $\Az \otimes \Az + \Cz$  is Schur stable. The proof of this property comes readily by noticing that such $\Nz$ corresponds to a descriptor representation of the system $\alpha_{0,k+1}^T=\alpha_{0,k}^T\left(\Az \otimes\Az +\Cz\right)$ and by following the lines of the proof of Theorem \ref{TH:stab_cond_A}, that system is stable if (\ref{LMI:specific}) holds.

According to these remarks, one natural {\it a priori} choice is $\Az=0$ and $\Cz=0$, providing a most Schur stable matrix $\Az \otimes\Az +\Cz=0$. Another case is, assuming some initial guess $K_0$ of $K$, to choose $\Az=A+BK_0$ and $\Cz=C_p^A$.

  
\subsection{The polytopic case}
By enforcing the lifted LMI in~(\ref{LMI:specific}) at the vertices $ A^{(i)}, B^{(i)}, \: \forall i=1,\ldots,L$, an extension of the result in Theorem~\ref{TH:stab_cond_A} is obtained for the case where $\theta$ is uncertain. The following corollary presents a sufficient condition for the Schur stability of the matrix $M(\theta,K), \: \forall \: \theta \in \mathbf{E}^{L}$. 

\begin{corollary}
Assume there exists $\Xii \in \mathbb{S}^{n^2}_+$, with $ i  =1,\ldots,L$, $\SM\in \R^{n\times n}$, $\T \in \R^{m \times n}$,  
and $\Nz \in \R^{3n^2\times 2n^2}$ 
such that 
the following holds for all $i=1,\ldots,L$ 
\begin{equation}
    \begin{pmatrix} 
    \Xii& 0 & 0  & \\
 0 & - \Xii & 0  & \\
  0 & 0 & 0  & 
\end{pmatrix} \prec
\left\{ N^{(i)}(\SM,\T) \NzT\right\}^{\mathcal{S}} 
\label{LMI:polytopic}
\end{equation}  
where $N^{(i)}(\SM,\T)=$
\[
\left(\begin{array}{cc|  cc }
- \SM \otimes I_n && 0  \\
C_p^A\left(\SM\otimes I_n \right) && I_n \otimes \left(A^{(i)} \SM+B^{(i)}\T\right) \\
\left(A^{(i)}\SM+B^{(i)}\T\right)\otimes I_n && -I_n\otimes \SM \end{array}\right)
\]
then $K = \T \SM^{-1}$ is such that $M(\theta,K)$ 
is robustly Schur stable $\forall \theta \in \mathbf{E}^{L}$. 
\end{corollary}

The proof is classical in the S-variable approach and amounts to notice that the inequalities (\ref{LMI:polytopic}) are affine in the matrices defining vertices of the polytope, hence by convexity of both the polytopes and the matrix inequalities, the conditions hold for all uncertainties.

\subsection{Links with Mean-Square Stability (MSS)}
Considering system~(\ref{Eq:sys_dyn_A}) without additive noise ($w_k = 0$), one can easily notice the link between the second-order moment of the state $\mathbb{E}\left[\lVert x_k \rVert^2\right]$ and the error covariance $\cov(e_k)$ since: 
    \begin{align}
     \mathbb{E}\left[\lVert x_k \rVert^2\right]&=  \mathbb{E} \left[\text{tr}\left(x_kx_k^T \right) \right] =\text{tr}\left( \mathbb{E}\left[x_kx_k^T\right]\right) \nonumber\\
     &= \text{tr}\left(\cov(x_k)\right)+\text{tr}\left( \mathbb{E}\left[x_k\right] \mathbb{E}\left[x_k\right]^T\right).    
    \label{Eq:trace_cov}
    \end{align}
Furthermore, since both $\cov(x_k)$ and $\mathbb{E}\left[x_k\right] \mathbb{E}\left[x_k\right]^T$ are positive semi-definite matrices, stabilizing the second-order moment ensures the stability of both the covariance and the expectation term $\mathbb{E}\left[x_k\right] \mathbb{E}\left[x_k\right]^T$. Moreover, one can notice that $\cov(x_k) = \cov(e_k)$ since by definition:
\begin{align}
\cov(x_k)&=\E\left[\left(x_k-\E\left[x_k\right]\right)\left(x_k-\E\left[x_k\right]\right)^T\right],\nonumber
\end{align}
by replacing $x_k=e_k+z_k$ and $\E\left[x_k\right]=z_k$, and since $\mathbb{E}\left[e_k\right]=0$, it follows directly that $\cov(x_k)=\cov(e_k)$.

Therefore, condition~(\ref{LMI:Yohei2019}) ensures the stability of the error covariance dynamics. Note, however, that proving the equivalence between the Schur stability of  $M(K)$ and condition~(\ref{LMI:Yohei2019}) needs further investigation. Indeed,  the Schur stability of $M(K)$ is sufficient, since it is established on the full vectorized space $\mathbb{R}^{n^2}$, whereas covariance matrices are symmetric and thus evolve on a lower-dimensional subspace. Finally, it is clear that both conditions imply the stability of the error covariance dynamics.


\section{Numerical results}\label{Section:Num_tests}
This section presents numerical comparisons and illustrations to assess the tractability and conservatism of the derived conditions.

\subsection{Tractability comparison}
We suggest to compare the computational complexity of the conditions in (\ref{LMI:specific}) and (\ref{LMI:Yohei2019}). Since the asymptotic stability notion is only related to systems without additive noise, we consider systems with $w_k=0$. Furthermore, in order to compare (\ref{LMI:specific}) to (\ref{LMI:Yohei2019}), we consider that only $\tilde{A}(\xi_k)$ is stochastic and the input matrix is deterministic. Note that in this case, the size of condition~(\ref{LMI:Yohei2019}) is further reduced using the singular value decomposition. 

The following setting has been considered: 

\begin{itemize}
    \item Random nominal matrices $A$ of size $n$ were generated such that both conditions were feasible and $B=I_n$. 
    \item The entries of $\bar{A}(\xi_k)$  are assumed to follow normal distributions $\mathcal{N}\left(0, 0.05\right)$.  
    \item The average computation time was evaluated using 100 tests, considering only the solver execution step. 
\end{itemize}

\begin{figure}
    \centering
\includegraphics[width=0.8\linewidth]{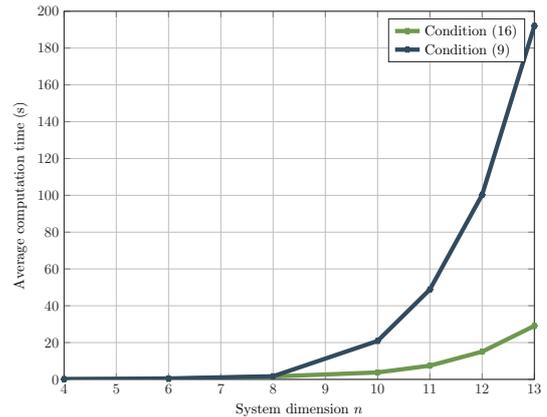}
    \caption{Average computation time over 100 tests}
    \label{fig:B_det}
\end{figure}

Fig.~\ref{fig:B_det} presents the average computation time for conditions~(\ref{LMI:specific}) and (\ref{LMI:Yohei2019}). The reported time corresponds only to the solver execution phase and does not include the decomposition step required for condition~(\ref{LMI:Yohei2019}). It can be observed that condition (\ref{LMI:specific}), whose size is $3n^2$, has a lower computational time compared to condition (\ref{LMI:Yohei2019}) with dimension reduction. Moreover, the difference in computation time increases with the increase of $n$.

\subsection{Conservatism assessment}
In order to evaluate the conservatism of the derived sufficient condition with respect to the necessary and sufficient one, let us consider systems as defined in~(\ref{Eq:sys_dyn_no_w}), where  the entries of $\bar{A}(\xi_k)$ follow the normal distribution defined as $\mathcal{N}\left(0, \sigma^2 \right)$ and $\tilde{B}(\xi_k)$  is considered deterministic that we denote $B=(1,0)^T$ for simplicity. For each setting, the variance is increased until the conditions are no more feasible, then, this maximal variance is denoted $\sigma_{max}^2$. 

Consider the following settings for system~(\ref{Eq:sys_dyn_no_w}) with a deterministic input matrix $B=(1,0)^T$:
\begin{align}
   \bm{\Gamma_1} \: :\:  A &= 
    \begin{pmatrix}
   3.5 &0.2\\
    1.2 & 1.5 
\end{pmatrix}, \: \text{mspec}(A) = \{             3.6136, 1.3864\}.  
  \nonumber
\end{align}
\begin{align}
   \bm{\Gamma_2} \: :\:  A &= 
    \begin{pmatrix}
   0.9 &0.1\\
    0.2 & 0.95 
\end{pmatrix}, \: \text{mspec}(A) = \{               0.7814,  1.0686\}. \nonumber
\end{align}
\begin{align}
   \bm{\Gamma_3} \: :\:  A &= 
    \begin{pmatrix}
   0.9 &0.1\\
    0.2 & 0.9 
\end{pmatrix}, \: \text{mspec}(A) = \{                1.0414,  0.7586\}. \nonumber 
\end{align}

\begin{table}
\centering
\renewcommand{\arraystretch}{1.5}
\arrayrulecolor{black}

\begin{tabular}{
    >{\columncolor{gray!25}}c |
    >{\columncolor{gray!25}}c >{\columncolor{gray!10}}c |
    >{\columncolor{gray!25}}c >{\columncolor{gray!10}}c
}
\rowcolor{gray!0}
 & \multicolumn{2}{c|}{\scriptsize \cellcolor{gray!50}Condition (\ref{LMI:specific})} & \multicolumn{2}{c}{\scriptsize \cellcolor{gray!50}Condition (\ref{LMI:Yohei2019})} \\
\rowcolor{gray!0}
 & \cellcolor{gray!25} \tiny{$\bm{\sigma^2_{max}}$} & \cellcolor{gray!25}\tiny{\bm{$K$}} & \cellcolor{gray!25}\tiny{\bm{$\sigma^2_{max}$}} & \cellcolor{gray!25}\tiny{\textbf{\bm{$K$}}} \\
\hline
\rowcolor{gray!10}
\scriptsize{\textbf{$\bm{\Gamma_1}$}} & -- & --
     & \tiny{0.09} &    \tiny \scalebox{1}{$                  \left(\hspace{1pt}\begin{matrix}
-4.6594  & -1.6551
        \end{matrix}\right)$} \\
\rowcolor{gray!10}
\scriptsize{\textbf{$\bm{\Gamma_2}$}} & \tiny{0.09} & \tiny \scalebox{1}{$     \left(\hspace{1pt}   \begin{matrix}
  -0.7888 &  -0.2967
        \end{matrix}\right)$} & \tiny{0.16} & \tiny \scalebox{1}{$ \left(\hspace{1pt}\begin{matrix}
-1.0372 &  -0.7751
        \end{matrix}\right)$} \\
\rowcolor{gray!10}
\scriptsize{\textbf{$\bm{\Gamma_3}$}} & \tiny{0.16} & \tiny  \scalebox{1}{$\left(\hspace{1pt}\begin{matrix} 
-0.9040 &  -0.3095
        \end{matrix}\right)$} & \tiny{0.2} & \tiny \scalebox{1}{$     \left(\hspace{1pt}   \begin{matrix}
 -1.0093  & -0.5969
        \end{matrix} \right)$} \\
\end{tabular}
\caption{Comparison of conditions (\ref{LMI:specific}) and (\ref{LMI:Yohei2019})}
\label{Table:A_stoc}
\end{table}
 
Table~\ref{Table:A_stoc} shows the maximal variance $\sigma_{max}^2$ for which the conditions in (\ref{LMI:specific}) and (\ref{LMI:Yohei2019}) are feasible, for different systems; in addition to the calculated pre-stabilizing gains $K$.  The symbol -- indicates infeasibility, even for very low variances. The presented results highlight the fact that the maximal variances for which the condition in  (\ref{LMI:specific}) remains feasible are lower than those related to condition (\ref{LMI:Yohei2019}). This difference in the maximal variances seems to increase with the instability of the nominal dynamics, measured by the deviation of the eigenvalues of $A$ from the unit circle.  This highlights the fact that the conservatism of the sufficient conditions in  (\ref{LMI:specific}) depends on the system under study. 

\subsection{Effects of the choice of $A_0$ and $C_0$}

In order to assess the effect of the choice of the auxiliary matrices \(A_0\) and \(C_0\) on the conservatism of the condition in~(\ref{LMI:specific}), 
we consider the setting denoted as \(\mathbf{\Gamma_3}\). 
We choose \(A_0 = A + BK\) with \(K = (-1.0093 \;\; -0.5969)\) resulting from the necessary and sufficient condition~(\ref{LMI:Yohei2019}), 
and \(C_0 = C_p^A\). 
With this choice of \(A_0\) and \(C_0\), the LMI in~(\ref{LMI:specific}) is feasible up to a variance of \(0.2\) and  the computed gain \(K = (-1.0091 \;\; -0.5982)\) is approximately the same as the one computed with condition~(\ref{LMI:Yohei2019}). 
This example is presented only to illustrate that, depending on the choice of the auxiliary matrices, 
the proposed sufficient condition can closely approximate the solution obtained from the necessary and sufficient one. 
Of course, it would be meaningless to assume a perfect choice of \(A_0\) and \(C_0\) in practice, but this motivates the use of heuristic approaches to this end.

\subsection{The polytopic case}

Consider system~(\ref{Eq:sys_dyn}) with the following setting:
\begin{align}
    A^{(1)} = 
    \begin{pmatrix}
   0.9 &0.2\\
    0.1 &0.8 
\end{pmatrix}, \: 
    A^{(2)} = 
    \begin{pmatrix}
   0.6 &0.2 \\
   0.1 &0.9
\end{pmatrix}, \:
    A^{(3)} = 
    \begin{pmatrix}
   0.9 &0.2 \\
    0.1 & 0.9
\end{pmatrix},
  \nonumber
\end{align}
\begin{align}
    A^{(4)} = 
    \begin{pmatrix}
   0.6 &0.2 \\
    0.1 & 0.8
\end{pmatrix}, \: 
    B^{(1,2,3,4)} = 
    \begin{pmatrix}
   1  \\
   0
\end{pmatrix}.  \nonumber
\end{align}
The entries of $\bar{A}(\xi_k)$ are assumed to follow a normal distribution $\mathcal{N}(0,0.15)$. We consider $A_0$ and $C_0$ as zeros matrices and we solve the LMI conditions in~(\ref{LMI:polytopic}), the resulting feedback gain is  \(K =   (-0.7783  \; \; -0.2162) \), and the average computational time over 100 solver executions is approximately 0.04~s.

\section{Conclusion}\label{Section:Conclusion}
This paper presented covariance-stabilization conditions for a class of stochastic systems, based on the S-variable approach. 
The derived conditions turn out to be numerically more tractable than existing necessary and sufficient ones, thanks to their reduced dimension.
Future work will focus on extending the class of considered stochastic systems, for instance to include systems with stochastic uncertainties in the input matrix, as well as relaxing the i.i.d.~assumption. 

\section*{Numerical setup}
All computations were performed on a MacBook Pro equipped with an Apple M4 Max chip (16-core CPU, 40-core GPU) and 64 GB of unified memory. The codes were implemented in MATLAB and the LMI conditions were solved using YALMIP \cite{Lofberg2004} and the SDPT3 solver \cite{SDPT32}.

\bibliography{ifacconf}             

\end{document}